\documentclass[twocolumn]{article}
\usepackage{epsfig}
\usepackage{graphicx}
\usepackage{color}
\usepackage[acmtitlespace,nocopyright]{acmneww}
\usepackage{times}
\usepackage{amsmath}
\usepackage{amssymb}
\usepackage{xspace}
\usepackage{url}

\newcommand {\mm}[1] {\ifmmode{#1}\else{\mbox{\(#1\)}}\fi}

\newcommand{\denselist}{\itemsep 0pt\parsep=1pt\partopsep 0pt}

\newcommand {\ignore}[1]{}

\newcommand{\proof}{\noindent{\sc Proof.~}}
\newcommand{\eop}{\hfill\usebox{\smallProofsym}\bigskip}  %
\newsavebox{\smallProofsym}                            % smallproofsymbol
\savebox{\smallProofsym}                               %
{%                                                     %
\begin{picture}(6,6)                                   %
\put(0,0){\framebox(6,6){}}                            %
\put(0,2){\framebox(4,4){}}                            %
\end{picture}                                          %
}                                                      %
\makeatletter
\long\def\@makecaption#1#2{%
  \vskip\abovecaptionskip
  \sbox\@tempboxa{\small #1: #2}%
  \ifdim \wd\@tempboxa >\hsize
    \small #1: #2\par
  \else
    \global \@minipagefalse
    \hb@xt@\hsize{\hfil\box\@tempboxa\hfil}%
  \fi
  \vskip\belowcaptionskip}
\makeatother

\newcommand{\uuu}           {\mm{\tt u}}

\newcommand{\UUU}           {\mm{\tt U}}
\newcommand{\AAA}           {\mm{\mathcal A}}

\newcommand{\RRR}           {\mm{\mathcal R}}

\newcommand{\voronoi}[2]    {\mm{\rm vor}_{#1}{({#2})}}
\newcommand{\restricted}[2] {\mm{\rm res}_{#1}{({#2})}}
\newcommand{\Restricted}[2] {\mm{\rm stack}_{#1}{({#2})}}

\newcommand{\Rspace}        {\mm{{\mathbb R}}}

\newcommand{\Edist}[2]      {\mm{\|{#1}-{#2}\|}}
\newcommand{\Ind}[1]        {\mm{\bf {#1}}}
\newcommand{\norm}[1]       {\mm{\|{#1}\|}}

\newcommand{\cupsp}         {{\; \cup \;}}

\newtheorem{result}{}

\title{The Medusa of Spatial Sorting: ~\\
       3D Kinetic Alpha Complexes and Implementation
       \thanks{This research is partially supported
               by NSF under grant DBI-0820624,
               by ESF under the Research Network Programme,
               and by the Russian Government under mega project 11.G34.31.0053.}
       }

\author{Michael Kerber\thanks{IST Austria (Institute of Science and
            Technology Austria), Kloster\-neu\-burg, Austria.} and
        Herbert Edelsbrunner\thanks{IST Austria (Institute of Science and
            Technology Austria), Kloster\-neu\-burg, Austria,
            Departments of Computer Science and of Mathematics,
            Duke University, Durham, North Carolina,
            and Geomagic, Research Triangle Park, North Carolina.}}

\begin{document}
\maketitle

\begin{abstract}
  Motivated by an application in cell biology, we consider spatial
  sorting processes defined by particles moving from an initial
  to a final configuration.
  We describe an algorithm for constructing a cell complex in space-time,
  called the \emph{medusa},
  that measures topological properties of the sorting process.
  The algorithm requires an extension of the kinetic data structures framework
  from Delaunay triangulations to fixed-radius alpha complexes.
  We report on several techniques to accelerate the computation.
\end{abstract}

\vspace{0.1in}

%%%%%%%%%%%%%%%%%%%%%%%%%%%%%%%%%%%%%%%%%%%%%%%%%%%%%%%%%%%%%%%%%%%%%%%%%%
%%%%%%%%%%%%%%%%%%%%%%%%%%%%%%%%%%%%%%%%%%%%%%%%%%%%%%%%%%%%%%%%%%%%%%%%%%
\section{Introduction}
\label{sec1}
%%%%%%%%%%%%%%%%%%%%%%%%%%%%%%%%%%%%%%%%%%%%%%%%%%%%%%%%%%%%%%%%%%%%%%%%%%
%%%%%%%%%%%%%%%%%%%%%%%%%%%%%%%%%%%%%%%%%%%%%%%%%%%%%%%%%%%%%%%%%%%%%%%%%%

Consider a finite set of particles or points in $\Rspace^3$,
moving in time along continuous trajectories.
Interpreting these points as the centers of moving objects,
we are interested in the topological changes the configuration undergoes.
Our interest in this problem originates in a sorting process that
segregates cells during zebrafish development,
as studied by Heisenberg and Krens \cite{HeKr11}.
The sorting process operates on intermixed configurations of cells in
which different types have different physical properties.
One example is a mix of two cell types, in which the cells of the first
type have a strong preference for neighboring cells of the same type
and a strong dislike of exposed boundary, while the cells of the
second type have milder preferences and dislikes.
The typical outcome is that the cells of the first type form a ball-like shape
that is engulfed by a spherical shell consisting of cells of the second type.

In an effort to formalize the sorting process and to make it amenable
to detailed and objective measurements, Heisenberg, Krens, and the authors
of this paper introduced the \emph{restricted Voronoi medusa} as a
mathematical representation.
It is a geometric body in $4$-dimensional space-time obtained by
stacking up restricted Voronoi regions in $\Rspace^3$ \cite{EHKK12}.
At any moment in time, the Voronoi region of a particle
is intersected with a ball, and the resulting bodies are glued together to
form a $4$-dimensional structure.
Applying persistent homology to the time function on this structure
yields fine-grained information about the sorting process
that is difficult to observe directly.
This paper complements this foundational work with a description
of the computational aspects of the medusa construction.

\paragraph{Results.}
\cite{EHKK12} proves that the restricted Voronoi medusa
has the same homotopy type as the medusa obtained by
stacking up simplices in the corresponding alpha complex.
The latter \emph{alpha medusa} is combinatorial in nature, which has
computational advantages.
Continuing in this direction, this paper makes three contributions:
\begin{enumerate}
  \item We describe an algorithm that maintains a fixed-radius alpha complex
    for points moving on piecewise algebraic trajectories in $\Rspace^3$.
    The algorithm supports insertions and deletions of points 
    and allows for piecewise algebraic trajectories.
  \item Maintaining an alpha complex, we construct the alpha medusa
    whose geometric and topological properties reflect the events during
    the sorting process.
  \item We convert the kinetic algorithm and the medusa construction
    into robust and efficient software.
    Basing the implementation on the \textsc{Cgal} package for
    kinetic data structures by Daniel Russel \cite{RuPhD07},
    it achieves correctness through the exact geometric
    computation paradigm.
\end{enumerate}
Part of Contribution 1 is an extension of previous work from kinetic
Delaunay triangulations to kinetic alpha complexes,
which is of independent interest.
The requirement of correctly comparing algebraic numbers,
without tolerance for inaccuracy or approximation in Contribution 3
seriously slows down the software, even for piecewise-linear trajectories.
To counteract, we introduce techniques that speed-up
the computations without sacrificing their correctness.
We evaluate the effectivity of these techniques experimentally.

\paragraph{Outline.}
Section \ref{sec2} explains background from computational geometry and topology.
Section \ref{sec3} describes the kinetic algorithm for
fixed-radius alpha complexes.
Section \ref{sec4} explains the algorithm
that constructs the medusa of a set of moving cells.
Section \ref{sec5} describes techniques to speed up the computations.
Section \ref{sec6} concludes the paper.

%%%%%%%%%%%%%%%%%%%%%%%%%%%%%%%%%%%%%%%%%%%%%%%%%%%%%%%%%%%%%%%%%%%%%%%%%%
%%%%%%%%%%%%%%%%%%%%%%%%%%%%%%%%%%%%%%%%%%%%%%%%%%%%%%%%%%%%%%%%%%%%%%%%%%
\section{Background}
\label{sec2}
%%%%%%%%%%%%%%%%%%%%%%%%%%%%%%%%%%%%%%%%%%%%%%%%%%%%%%%%%%%%%%%%%%%%%%%%%%
%%%%%%%%%%%%%%%%%%%%%%%%%%%%%%%%%%%%%%%%%%%%%%%%%%%%%%%%%%%%%%%%%%%%%%%%%%

We review the fundamental geometric data structures that are required
in this work.
Voronoi tessellations and Delaunay triangulations are treated
in most computational geometry textbooks, including \cite{dBCKO08,Ede01},
and alpha complexes are described in \cite{EdHa10,EdMu94}.
For the most part, the discussion focuses on the $3$-dimensional case.
Most definitions and properties extend to higher dimensions
as well as to the plane.
We will exploit the latter fact when we draw illustrations.

\paragraph{Cell complexes.}
We recall that a \emph{$k$-simplex} is the convex hull
of an affinely independent set of $k+1$ points in some Euclidean space.  
A \emph{face} is a simplex defined by a subset of the $k+1$ points.
It is \emph{proper} if the subset is different from the set.
Reversing the direction, we call the $k$-simplex a \emph{coface} of its face.
We define a \emph{simplicial complex} as a finite collection of simplices
that is closed under the face relation, with the additional property
that any two simplices in the collection are either disjoint or
their intersection is a face of both.
The \emph{boundary} of a $k$-simplex is the collection of its $(k-1)$-faces.
The simplices of dimension $0$, $1$, $2$, and $3$ are referred to as
\emph{vertices}, \emph{edges}, \emph{triangles}, and \emph{tetrahedra}.
The \emph{star} of a $k$-simplex is the set of simplices that contain
the $k$-simplex as a face.
Noting that the star is in general not closed under the face relation,
we define the \emph{closed star} as the set of all simplices
in the star and of all faces of these simplices.
It is the smallest simplicial complex that contains the star.
Finally, if $\sigma$ is a $k$-simplex and $u$ is a point that does not
lie in the $k$-plane of the simplex, then the \emph{join}, denoted as
$u \ast \sigma$, is the $(k+1)$-simplex that is the convex hull
of $u$ and the vertices of $\sigma$.

It is convenient to also introduce an abstract counterpart
to the above geometric concept of a simplicial complex.
Specifically, an \emph{abstract simplicial complex} consists of a
finite set of (abstract) elements and a collection of subsets that
is closed under the subset relation.
We may map each element to a point in some Euclidean space, and each
subset to the convex hull of the points that correspond to its elements.
If the dimension of the space is sufficiently high and
the points are well chosen, this is a simplicial complex,
which we refer to as a \emph{geometric realization} of the abstract
simplicial complex.
Here is an example of this construction.
Consider a finite set, $X$, of possibly overlapping bodies,
and define the \emph{nerve} as the collection of subsets of $X$
with non-empty common intersection.
We note that the nerve is an abstract simplicial complex.
Indeed, the bodies are the elements, and if $A \subseteq X$
is a set in the nerve, then every subset of $A$ is also in the nerve.
A useful result is the Nerve Theorem \cite{EdHa10},
which states that if the bodies in $X$ are convex then
every geometric realization of the nerve has the same homotopy type
as the union of the bodies.
Intuitively, this means that one can be transformed into the other
by continuous transformations like bending, shrinking, and expanding,
but without gluing and cutting.
Most complexes in this paper will be constructed using the nerve operation.

We will have occasion to also use complexes that are more general
than simplicial complexes.
Instead of $k$-simplices, they contain \emph{$k$-cells}, which are
homeomorphic to the $k$-dimensional unit ball.
The \emph{boundary} of a $k$-cell is the homeomorphic image
of the $(k-1)$-sphere that bounds the $k$-ball.
We define a \emph{cell complex} as a finite collection of cells
with pairwise disjoint interiors such that the boundary of each
$k$-cell is a union of $(k-1)$-cells in the complex.

\paragraph{Voronoi tessellations and Delaunay complexes.}
Consider now a finite set of points, $U$, in $\Rspace^3$.
The \emph{Voronoi region} of a point $u$ in $U$ is the set of points
$x \in \Rspace^3$ that have $u$ as the closest point in $U$:
\begin{eqnarray*}
  \voronoi{U}{u} &=& \{ x\in\Rspace^3 \mid \Edist{x}{u}\leq \Edist{x}{v},
                       \forall v \in U \}.
\end{eqnarray*}
Note that $\voronoi{U}{u}$ is convex.
We usually drop $U$ from the notation.
The \emph{Voronoi tessellation} of $U$ is the set of Voronoi regions
of its points.
While it is not a cell complex formally, we get a cell complex
if we add the common intersections of Voronoi regions as
lower-dimensional cells to the set.
If the points in $U$ are in general position, by which we mean that no
four lie in a common plane and no five lie on a common sphere,
then the Voronoi regions intersect in a rather predictable pattern.
Specifically, the intersection of any two is either empty or a
($2$-dimensional) polygon, the intersection of any three is either empty
or a ($1$-dimensional) edge, and the intersection any four is either
empty or a ($0$-dimensional) point.
Furthermore, the intersection of five or more Voronoi regions
is necessarily empty.

We get the dual \emph{Delaunay complex} if we replace each
non-empty intersection of Voronoi regions by the convex hull
of the points that generate the Voronoi regions containing the intersection.
Equivalently, we may define the Delaunay complex as the set
of convex hulls of subsets of points that have the \emph{empty sphere property}.
Specifically, this means that there exists a sphere that passes through
the points of the subset and all other points in $U$
lie strictly outside this sphere.
We note that the center of this sphere belongs to the intersection
of the corresponding Voronoi regions.
Assuming general position, the Delaunay complex is a simplicial complex,
which is generally referred to as the \emph{Delaunay triangulation}.
It is a geometric realization of the nerve of the Voronoi tessellation.

\paragraph{Restricted Voronoi tessellations and alpha complexes.}
Fixing a positive radius, $\alpha_0$, we define
the \emph{restriction} of a Voronoi region to be its intersection
with the closed ball of radius $\alpha_0$ centered at the generating point:
\begin{eqnarray*}
  \restricted{U}{u}  &=&  \{ x \in \voronoi{U}{u} \mid
                             \Edist{x}{u} \leq \alpha_0 \}.
\end{eqnarray*}
Again, we usually drop $U$ from the notation.
The \emph{restricted Voronoi tessellation} of $U$ is the set of
restricted Voronoi regions of its points.
In contrast to the unrestricted case, each restricted Voronoi region
is bounded, and the tessellation covers only the union of balls
and not the entire space.
  
As before, we assume general position so we can dualize
by geometrically realizing the nerve.
The resulting simplicial complex is called the \emph{alpha complex}.
Since $\restricted{}{u} \subseteq \voronoi{}{u}$, for each point $u$ in $U$,
the alpha complex is a subcomplex of the Delaunay triangulation.
Next, we derive an equivalent condition for a Delaunay simplex
to lie in the alpha complex which is more suitable for computations.
Each $k$-simplex in the Delaunay triangulation
has a unique circumscribed $(k-1)$-sphere in its supporting $k$-plane.
We call its center the \emph{circumcenter}, its radius the \emph{circumradius},
the ball in $\Rspace^3$ with this center and this radius the \emph{circumball},
and the sphere that bounds the circumball
the \emph{circumsphere} of the $k$-simplex.
Note that the circumsphere is the smallest sphere that passes through the
vertices of the $k$-simplex.
We call the $k$-simplex \emph{short} if its circumradius is smaller than
or equal to $\alpha_0$.
Finally, we call the $k$-simplex \emph{Gabriel} if
its circumball has no point of $U$ in its interior.
\begin{result}[Short\&Gabriel Lemma]
  A simplex in the Delaunay triangulation of $U$ belongs
  to the alpha complex, for radius $\alpha_0$,
  iff it is short and Gabriel,
  or it is the face of another Delaunay simplex
  that is short and Gabriel.
\end{result}
The face of a short simplex is necessarily short,
but the face of a Gabriel simplex is not necessarily Gabriel.
It follows that all simplices in the alpha complex are short,
but not all simplices need to be Gabriel.
Also note that a tetrahedron is in the Delaunay triangulation iff
it is Gabriel; therefore, it is in the alpha complex iff it is short.
In our application, we use the restricted Voronoi tessellation
to model a set of biological cells for which the positions
of their nuclei are known.
Indeed, a cell tends to minimize its surface area and usually
does not grow larger than a certain size.
Therefore, a restricted Voronoi region appears to be a good approximation
of the actual cell shape and is still simple enough
for our computational purposes.

%%%%%%%%%%%%%%%%%%%%%%%%%%%%%%%%%%%%%%%%%%%%%%%%%%%%%%%%%%%%%%%%%%%%%%%%%%
%%%%%%%%%%%%%%%%%%%%%%%%%%%%%%%%%%%%%%%%%%%%%%%%%%%%%%%%%%%%%%%%%%%%%%%%%%
\section{Kinetic Alpha Complexes}
\label{sec3}
%%%%%%%%%%%%%%%%%%%%%%%%%%%%%%%%%%%%%%%%%%%%%%%%%%%%%%%%%%%%%%%%%%%%%%%%%%
%%%%%%%%%%%%%%%%%%%%%%%%%%%%%%%%%%%%%%%%%%%%%%%%%%%%%%%%%%%%%%%%%%%%%%%%%%

In this section, we describe the algorithm that maintains the alpha
complex for a fixed radius $\alpha_0 > 0$.
We pay particular attention to the certificates that govern
the sequence of operations needed to preserve the correctness of
the structure at all times.

\paragraph{The kinetic framework.}
The input to our algorithm is a finite set of trajectories,
each a continuous map $\uuu: [a,b] \to \Rspace^3$ with $0 \leq a < b \leq 1$.
For simplicity, we assume it to be piecewise linear,
with $a = t_0 < t_1 < \ldots < t_k = b$ such that
there are points $a_j, b_j \in \Rspace^3$ for which
$\uuu(t) = (1-t) a_j + t b_j$ for $t_j \leq t \leq t_{j+1}$.
In other words, we can write $\uuu(t) = (f_1(t), f_2(t), f_3(t))$
such that between $t_j$ and $t_{j+1}$,
each $f_i$ is a polynomial of degree $1$.
We call $t_0, t_1, \ldots, t_k$ the \emph{bending events} of the trajectory.
Furthermore, we assume that the trajectories do not meet each other, that is,
$\uuu(t) \neq \uuu'(t)$ for all $\uuu, \uuu' \in \UUU$
and all $t$ for which both trajectories are defined.

Our task is to maintain a data structure that 
goes from an initial configuration, at time $t=0$,
to the final configuration, at time $t=1$.
For that, the data structure is constructed at time $t=0$,
and maintained through a sequence of update operations
until the final configuration is reached.
It is assumed that the number of updates is finite,
and we call the time of an update an \emph{event}.
Events are detected by defining suitable \emph{certificate functions},
also referred to as \emph{certificates}. 
At any moment $t$ different from any event,
we have a collection of \emph{active} certificates, all being non-zero at $t$.
Importantly, they guarantee that as long as no certificate
changes its sign, our data structure remains structurally unchanged.
To detect the next event, the algorithm then finds the smallest root
of any active certificate that is greater than $t$.
It handles the event by updating the data structure
and the collection of active certificates.
Throughout this paper, we make the simplifying assumption
that all events are distinct, that is,
no two events happen at the same moment in time;
see also Section \ref{sec6}.

\begin{figure}[hbt]
 \centering
  \includegraphics[width=7cm]{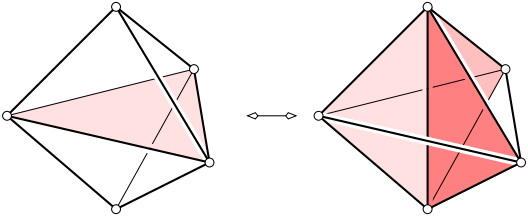}
  \caption{Illustration of a $2$-$3$-flip that alters the triangulation
     of a triangular double pyramid.
     On the left, the five points span two tetrahedra
     meeting in a triangle.
     After the flip, the triangle is replaced by an edge
     and the three incident triangles that connect
     the edge to the remaining three points.}
 \label{fig:2-3-flip}
\end{figure}

\paragraph{Maintaining the Delaunay triangulation.}
Since we need it later, we begin by reviewing the kinetic algorithm
for $3$-dimensional Delaunay triangulations
described in \cite{RuPhD07}.
Besides changes brought about by insertions and deletions of points,
and switches to new trajectory segments, there are only two configurations
that trigger a structural change in the triangulation:
\begin{itemize}
  \item five points of $U$ lie on a common sphere, and the open
    ball bounded by this sphere contains no points of $U$;
  \item four points of $U$ lie on a common plane, and one of the
    open half-spaces bounded by this plane contains no points of $U$.
\end{itemize}
We call each such configuration a \emph{degeneracy}.
Consistent with the above assumption of distinct events,
we assume that at every moment of time there is only one degeneracy,
and that each degeneracy lasts only for a single moment.
In other words, we can find a small open interval in time
during which the given degeneracy exists at a single point in time,
and it is the only degeneracy that occurs during this interval.
We can therefore study the effect of the degeneracy by considering
the non-degenerate local configurations right before and right after
the degeneracy.
Consider for example a degeneracy of the first type,
which involves five points.
Right before the degeneracy, the five points span two Delaunay tetrahedra
with a common triangle,
and right after the degeneracy they span three tetrahedra
so that each pair shares a triangle and all three share an edge.
Of course, it can also be the other way round.
Importantly, we can transform one configuration to the other by flipping.
In this particular case, we substitute three for two or two for three
tetrahedra, calling the operation a \emph{$2$-$3$-flip};
see Figure \ref{fig:2-3-flip}.
To avoid a case analysis, we represent the triangulation using a
vertex at infinity that is joined to every simplex
in the boundary of the convex hull of $U$.
Effectively, we embed the triangulation on a $3$-sphere.
This way, we can add the point at infinity to the set of four points
forming a degeneracy of the second type, thus getting a degeneracy of
the first type, which is handled by a $2$-$3$-flip, as described above.

\paragraph{Flip events.}
The transition of the Delaunay triangulation across
degenerate configurations is controlled by two certificate functions.
Let $\uuu^\Ind{1}, \uuu^\Ind{2}, \uuu^\Ind{3}, \uuu^\Ind{4}, \uuu^\Ind{5}$
be the five trajectories of the points that span two tetrahedra
sharing a triangle or three tetrahedra sharing a common edge,
as in Figure \ref{fig:2-3-flip}.
If one of the trajectories belongs to the infinite vertex then we
reorder them such that this trajectory is $\uuu^\Ind{5}$.
Let $\uuu^\Ind{i} (t) = (f^\Ind{i}_1 (t), f^\Ind{i}_2 (t), f^\Ind{i}_3 (t))$ 
be the coordinate functions of the finite points, and
recall that the squared norm of this point
is the sum of the squares of its three coordinates.
If all five points are finite, we create the certificate
\begin{eqnarray}
  \det \left[ \begin{array}{ccccc} 
                1 & f^\Ind{1}_1 (t) & f^\Ind{1}_2 (t) & f^\Ind{1}_3 (t) & \norm{\uuu^\Ind{1} (t)}^2 \\
                1 & f^\Ind{2}_1 (t) & f^\Ind{2}_2 (t) & f^\Ind{2}_3 (t) & \norm{\uuu^\Ind{2} (t)}^2 \\
                1 & f^\Ind{3}_1 (t) & f^\Ind{3}_2 (t) & f^\Ind{3}_3 (t) & \norm{\uuu^\Ind{3} (t)}^2 \\
                1 & f^\Ind{4}_1 (t) & f^\Ind{4}_2 (t) & f^\Ind{4}_3 (t) & \norm{\uuu^\Ind{4} (t)}^2 \\
                1 & f^\Ind{5}_1 (t) & f^\Ind{5}_2 (t) & f^\Ind{5}_3 (t) & \norm{\uuu^\Ind{5} (t)}^2
             \end{array} \right] ,
  \label{eqn:flipA}
\end{eqnarray}
which is a univariate polynomial in $t$
that is zero iff the five points are co-spherical.
Assuming the coordinate functions are linear,
the degree of the polynomial is $5$.
If the fifth point is at infinity, we create the certificate
\begin{eqnarray}
  \det \left[ \begin{array}{cccc} 
               1 & f^\Ind{1}_1 (t) & f^\Ind{1}_2 (t) & f^\Ind{1}_3 (t) \\
               1 & f^\Ind{2}_1 (t) & f^\Ind{2}_2 (t) & f^\Ind{2}_3 (t) \\
               1 & f^\Ind{3}_1 (t) & f^\Ind{3}_2 (t) & f^\Ind{3}_3 (t) \\
               1 & f^\Ind{4}_1 (t) & f^\Ind{4}_2 (t) & f^\Ind{4}_3 (t)
             \end{array} \right] ,
  \label{eqn:flipB}
\end{eqnarray}
which is zero iff the four point are coplanar.
We call the polynomials in (\ref{eqn:flipA},\ref{eqn:flipB})
\emph{flip certificates} and their roots \emph{flip events}.

After having constructed the initial certificates, at time $t = 0$, 
the algorithm finds the first positive flip event.
It then performs a $2$-$3$-flip, creating certificates for the
(new) simplices inside the double pyramid, and updating the certificates
of the simplices in the boundary of the double pyramid.
The updating is necessary because the star of each boundary simplex
changes during the flip.
After these steps, both data structure and certificates are again valid,
and the iteration continues with the next flip event.

\paragraph{Radius events.}
Next, we extend the kinetic algorithm from Delaunay triangulations
to alpha complexes.
As before, we use a fixed radius $\alpha_0 > 0$.
We represent the alpha complex by equipping each Delaunay simplex
with a flag that indicates whether or not the simplex
belongs to the alpha complex.
To construct these flags at time $t = 0$,
we check every Delaunay simplex for being short and for being Gabriel.
Following the Short\&Gabriel Lemma in Section \ref{sec2},
we add all Delaunay simplices that are short and Gabriel,
as well as all their faces, to the alpha complex.
To maintain the flags, we construct a certificate for each edge, triangle,
and tetrahedron whose roots are the times when
the circumradius of the simplex equals $\alpha_0$.
To simplify the discussion, we assume the generic case in which
the circumradius changes from strictly smaller to strictly larger
than $\alpha_0$, or vice versa.
We call these functions \emph{radius certificates}
and their roots \emph{radius events}.
Whenever a Delaunay simplex is inserted or deleted,
the algorithm also creates or removes the corresponding radius certificate.
The certificate of an edge compares the length to $2 \alpha_0$,
and taking squares, we get a polynomial of degree $2$.
The radius certificates of a triangle and a tetrahedron are more complicated,
but can be derived from suitable minors of the matrix
that defines the circumsphere of the simplex;
see \cite{WeiWeb} for the formula in the tetrahedral case.
We will discuss the triangle case in Section \ref{sec5}.  

After initializing the alpha complex and the certificates,
the algorithm looks for the next event.
If this is a flip event, we proceed as described above.
In addition, we update the flags that identify the alpha complex
as a subcomplex of the Delaunay triangulation.
Because all tetrahedra involved in the $2$-$3$-flip
have the same circumsphere, they are either all short or all non-short.
If they are short, all new Delaunay simplices are added to the alpha complex,
and otherwise, none of them is added to the alpha complex.
Second, consider the case in which the next event is a radius event.
Let $\sigma$ be the corresponding Delaunay simplex.
If $\sigma$ goes from non-short to short,
then its proper cofaces are necessarily non-short.
We check whether $\sigma$ is also Gabriel,
noting that this is always the case if $\sigma$ is a tetrahedron.
If so, $\sigma$ is added to the alpha complex together will all its faces.
On the other hand, if $\sigma$ goes from short to non-short,
then its proper faces are necessarily short.
We remove $\sigma$ from the alpha complex, unless is was not in
the complex even before the event.
If the event causes the deletion of $\sigma$ from the alpha complex,
then this may have consequences for its faces.
In particular, if $\sigma$ was the last proper coface of $\tau$
in the alpha complex, and $\tau$ is not Gabriel,
then $\tau$ is also deleted from the alpha complex. 
Afterwards, the algorithm continues with the next event.

\paragraph{Redundancy of Gabriel events.}
Perhaps surprisingly, flip and radius events suffice to maintain
the alpha complex.
Flip certificates monitor when tetrahedra become non-Delaunay,
and radius events monitor when simplices become short or non-short.
We do not need certificates that monitor when simplices become
Gabriel or non-Gabriel.
To understand why such certificates are not necessary, we
call a time $t$ \emph{G-critical} for a simplex $\sigma$,
if at that time, $\sigma$ changes from Gabriel to non-Gabriel, or vice versa.
\begin{result}[G-criticality Lemma]
  Let $t$ be a G-critical time for a short Delaunay edge or triangle.
  Then this edge or triangle has a proper coface that is in
  the alpha complex at time $t$.
\end{result}
\proof
 Denote the edge or triangle by $\sigma$ and consider its circumball,
 $B_\sigma$, at time $t$.
 No point of $U$ lies in the interior of $B_\sigma$,
 but there is a point $u$ on the bounding sphere
 that is not a vertex of $\sigma$.
 The join $u \ast \sigma$ is another simplex in the Delaunay triangulation,
 and it is a proper coface of $\sigma$.
 It has the same circumball as $\sigma$,
 which implies that $u \ast \sigma$ is short and Gabriel and
 therefore belongs to the alpha complex at time $t$.
\eop

The lemma implies that when a short edge or triangle changes
its Gabriel status, it is a face of a simplex in the alpha complex.
The status change has therefore no impact on its membership
in the alpha complex.

\paragraph{Other events.}
For later reference, we briefly mention the remaining types of events
supported by our algorithm.
First, we consider a bending event, at which a trajectory
starts a new segment.
Such an event leaves the alpha complex unchanged, 
but all flip and radius certificates that involve the coordinates
of the corresponding vertex are recomputed.
These are the certificates associated to the simplices in the closed star
of the vertex.

Second, we allow for insertions and deletions of points.
The two operations are mostly symmetric, and we only discuss
the insertion of a point $u$.
We add $u$ to the Delaunay triangulation by identifying all tetrahedra
whose circumballs contain $u$,
referring to their union as the \emph{conflict region} of $u$.
Since these tetrahedra no longer satisfy the empty sphere criterion,
we remove them from the Delaunay triangulation,
together with their faces in the interior of the conflict region.
Next, we add $u$ and connect it to all simplices in the boundary
of the conflict region.
After the operation, these simplices form the boundary of
the closed star of $u$.
Finally, the simplices in the closed star are checked for being in
the alpha complex, and their certificates are created or updated.

%%%%%%%%%%%%%%%%%%%%%%%%%%%%%%%%%%%%%%%%%%%%%%%%%%%%%%%%%%%%%%%%%%%%%%%%%%
%%%%%%%%%%%%%%%%%%%%%%%%%%%%%%%%%%%%%%%%%%%%%%%%%%%%%%%%%%%%%%%%%%%%%%%%%%
\section{Medusa Construction}
\label{sec4}
%%%%%%%%%%%%%%%%%%%%%%%%%%%%%%%%%%%%%%%%%%%%%%%%%%%%%%%%%%%%%%%%%%%%%%%%%%
%%%%%%%%%%%%%%%%%%%%%%%%%%%%%%%%%%%%%%%%%%%%%%%%%%%%%%%%%%%%%%%%%%%%%%%%%%

In this section, we focus on the construction of space-time complexes
that give a static $4$-dimensional representation of a dynamic
spatial process.
After a brief review of the medusa concept,
we discuss its construction using kinetic algorithms.

\begin{figure}[thb]
 \centering
   \includegraphics[width=7.0cm]{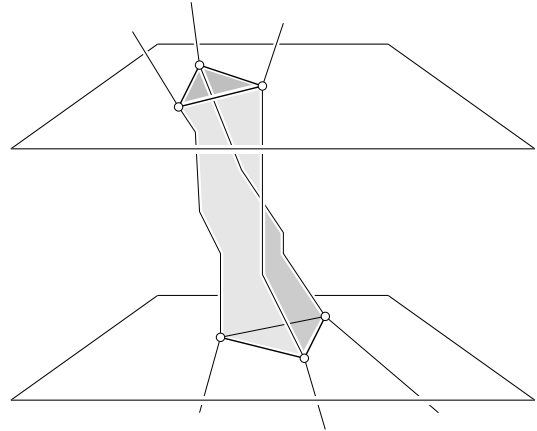}
   \caption{The triangle spanned by points on the three trajectories
     sweeps out a triangular prism while it
     belongs to the alpha complex between the lower and the upper planes.}
 \label{fig:alpha_medusa_illu}
\end{figure}

\paragraph{Restricted Voronoi and alpha medusa.}
Let $\UUU$ be a finite set of trajectories,
and write $\UUU(t) \subseteq \Rspace^3$ for the set of points obtained
by evaluating all trajectories at time $t$.
Of course, we evaluate only those trajectories that are defined at $t$.
The restricted Voronoi tessellation of $\UUU(t)$ is constructed
as described in Section \ref{sec2}.
Letting $\uuu: [a,b] \to \Rspace^3$ be a particular trajectory,
we get the region of $\uuu(t)$ within the tessellation of $\UUU(t)$
for each time $a \leq t \leq b$.
Piling up these regions in $\Rspace^4$ and taking the closure, 
we call the result
the \emph{stack} of the trajectory $\uuu$:
\begin{eqnarray*}
  \Restricted{}{\uuu}  &=&  
     \mathrm{clos}\left(\bigcup_{a \leq t \leq b}
       [ \restricted{\UUU(t)}{\uuu(t)} \times t ] \right).
\end{eqnarray*}
Taking the closure is not redundant because the region of a trajectory can
change discontinuously when inserting another trajectories, causing
a locally open pile.
The set of stacks of trajectories in $\UUU$ is
the \emph{restricted Voronoi medusa}, which we denote as $\RRR = \RRR(\UUU)$.
While it is the preferred representation of a spatial sorting process,
we introduce a dual structure that is easier to compute
and gives the same measurements.
As a first approximation of this construction,
we exploit the duality between restricted Voronoi diagrams and alpha complexes
and pile up the simplices of the latter to form cells in $\Rspace^4$;
see Figure \ref{fig:alpha_medusa_illu}.
This construction has two shortcomings.
First, the cells do not define a cell complex,
but this can be corrected by adding (geometrically degenerate) cells
that represent flip, insertion, and deletion events.
Second, the walls are redundant and are better contracted.
The contraction produces simplices, but they
are connected to each other in more general ways
than allowed in a simplicial complex.

The main difficulty in defining the alpha medusa formally is
a lack of regularity of the stacks in $\RRR$.
While every restricted Voronoi region is convex,
the stack of such regions is generally not convex,
and the intersection of two stacks is generally not connected.
Taking the nerve of the set of stacks would thus produce an
abstract complex that does not respect the Nerve Theorem.
To finesse these difficulties,
we take the \emph{multi-nerve} as follows.
If $k+1$ stacks intersect in $\ell$ different connected components,
then the multi-nerve has $\ell$ copies of the abstract $k$-simplex,
each representing one component of the intersection. 
Moreover, each abstract simplex is associated with
the \emph{lifetime interval}, $[t_1,t_2]$, of the corresponding
connected component.
Because there is at most one component at every moment in time, 
we have disjoint intervals for different copies of the same simplex.
The case $t_1 = t_2$ is allowed and represents
connectivity at a single moment in time.
The result of this multi-nerve construction is what we call
the \emph{alpha medusa} and denote as $\AAA = \AAA(\UUU)$.
In \cite[Lemma B]{EHKK12}, it is shown that $\AAA$
is homotopy equivalent to $\RRR$ as well as the space
covered by the prismatic cells swept out by simplices of the alpha complex.

\paragraph{Construction.}
A simplex with lifetime interval $[t_1, t_2]$ is \emph{active}
for all $t_1 \leq t < t_2$ and \emph{finished} for all $t_2 \leq t$.
We construct the alpha medusa by running the kinetic alpha complex algorithm
and updating the medusa at each event.
We maintain two lists of simplices,
called the \emph{output list} and the \emph{active list},
preserving the following property at all times:
\begin{result}[Invariant]
  After the algorithm has handled an event at time $t$,
  the output list consists of all simplices finished at $t$,
  and the active list consists of all simplices active at $t$.
\end{result}
At the beginning, all simplices in the alpha complex
are put into the active list.
Whenever a simplex is added to the alpha complex at a time $t$, 
we put it into the active list,
and when it is removed from the alpha complex at a later time $t'$,
we remove it from the active list and put it into the output list
with lifetime interval $[t,t']$.

There are additional steps to be taken during special types of events.
We discuss flip events now and the more complicated insertion and
deletion events later.
Consider a $2$-$3$-flip at time $t$. 
We have two tetrahedra before and three after the flip, or vice versa.
At time $t$, they all have the same circumball,
so either all or none of them belong to the alpha complex.
In the former case, the five corresponding restricted Voronoi regions
meet in the center of the common circumball.
It follows that the dual $4$-simplex belongs to the multi-nerve,
with lifetime interval $[t, t]$.
We thus add the $4$-simplex, connecting it its five boundary tetrahedra,
which are the ones involved in the flip.
Indeed, the $4$-simplex fills the void between the five tetrahedra,
which could not be filled by gluing the tetrahedra to each other
as they are not face-to-face.

\paragraph{Insertions and deletions.}
Finally, we consider the case in which a trajectory, $\uuu'$,
is inserted at time $t$.
The case of a deletion is symmetric and omitted.
Let $u' = \uuu'(t)$ be the vertex inserted into the alpha complex.
As described in Section \ref{sec3}, the insertion of $u'$
into the Delaunay triangulation is synonymous with the substitution
of the star for the simplices in the conflict region of $u'$.
Any subset of these simplices may belong to the alpha complex.
In addition to finding this subset, we need to construct simplices
that fill the gap between the deleted and the inserted alpha complex
simplices in the medusa.
This is similar to a flip,
in which the void formed by the five tetrahedra is filled by a $4$-simplex.
To describe a general solution to this problem,
let $U$ be a finite set of points, $u'$ a point not in $U$,
and write $U' = U \cupsp \{u'\}$ for the set including the new point.
We call each $\restricted{U}{u}$ an \emph{old region},
each $\restricted{U'}{u}$ a \emph{new region},
and $\restricted{U'}{u'}$ the \emph{region} of $u'$.
\begin{result}[Insertion Lemma]
  Let $I$ be a non-empty common intersection of old regions.
  Then, $I$ intersects the region of $u'$ iff
  the common intersection of the corresponding
  new regions is either empty or it also intersects the region of $u'$.
\end{result}
\proof
  For the forward direction, assume that $I$ intersects the region of $u'$,
  and let $x$ be an arbitrary point in that intersection.
  Note that $\Edist{x}{u'} \leq d(x)$,
  where $d(x)$ is the distance to the closest points in $U$.
  In the first case, we have $\Edist{y}{u'} < d(y)$, for all $y \in I$,
  which implies that the corresponding new regions do not intersect.
  In the second case, we have $\Edist{y}{u'}\geq d(y)$, for some $y$,
  Then there exists a point $y' \in I$ such that $\Edist{y'}{u'} = d(y')$,
  which implies that $y'$ belongs to the intersection of the new regions
  with the region of $u'$.
  
  For the backward direction, if the 
  new regions intersect the region of $u'$, so do the old regions. 
  In the remaining case, the new regions do not intersect.
  Let $x$ be a point in the intersection of the old regions.
  Since $x$ is not in the intersection of the new regions,
  it must be in the region of $u'$. 
\eop
 
We use the Insertion Lemma to derive an algorithm that updates the
alpha medusa upon the insertion of a point $u'$ at time $t$.
Similar to the case of the Delaunay triangulation, we identify
the conflict region and remove all simplices in its interior.
In addition, if a removed $k$-simplex $\sigma$ is in the alpha complex
just before $t$, we add the join, $u' \ast \sigma$,
to the alpha medusa with lifetime interval $[t,t]$.
We do this because the $k+1$ old regions represented by $\sigma$
have a non-empty common intersection, but the corresponding new regions do not.
In the second step, we insert the star of $u'$ into the Delaunay triangulation,
check for every inserted simplex whether it belongs to the alpha complex,
and if it does then we add it to the active list.
Clearly, every inserted simplex is of the form $u' \ast \sigma$,
with $\sigma$ on the boundary of the conflict region.

%%%%%%%%%%%%%%%%%%%%%%%%%%%%%%%%%%%%%%%%%%%%%%%%%%%%%%%%%%%%%%%%%%%%%%%%%%
%%%%%%%%%%%%%%%%%%%%%%%%%%%%%%%%%%%%%%%%%%%%%%%%%%%%%%%%%%%%%%%%%%%%%%%%%%
\section{Implementation and Experiments}
\label{sec5}
%%%%%%%%%%%%%%%%%%%%%%%%%%%%%%%%%%%%%%%%%%%%%%%%%%%%%%%%%%%%%%%%%%%%%%%%%%
%%%%%%%%%%%%%%%%%%%%%%%%%%%%%%%%%%%%%%%%%%%%%%%%%%%%%%%%%%%%%%%%%%%%%%%%%%

In this section, we turn to implementation issues.
In particular, we discuss how to implement the algorithm in a robust manner,
we study the effect of practical choices,
and we present experimental results obtained with our software.

\paragraph{Robust computation.}
Recall the basic structure of a kinetic data structure as explained
in Section \ref{sec3}:
it consists of certificate functions, which are polynomials in $t$,
and each step advances the state to the smallest root larger than the
current time.
To maintain the certificates and to advance to the next event,
the algorithm computes and compares real roots of univariate polynomials.
These roots are algebraic numbers --- irrational in general ---
which makes the computations non-trivial.
The naive solution of approximating these roots by inexact
floating-point numbers can have unpredictable effects.
It is not true that the outcome is just slightly wrong,
e.g.\ by switching the order of events that happen almost simultaneously,
but the incorrect order can lead to inconsistent configurations,
causing program crashes, non-termination, and non-sensical results.
This problem is well-known in geometric contexts \cite{KMPSY08} and several 
approaches have been proposed.
We follow the \emph{exact geometric computation (EGC) paradigm},
popularized by Chee Yap \cite{Yap97}.
It suggests that the basic primitives be mathematically correct,
so that an algorithm using these primitives is in the position
to compute provably correct results.
Translated to our situation, we require that the events of our process
are handled in the mathematically correct order. 
The price we pay for this interpretation of robustness
is the burden to compute with algebraic numbers.
 
We implement our algorithm using the
\textsc{Cgal} library\footnote{Computational Geometry Algorithms Library,
                               \texttt{www.cgal.org}.},
which is designed in the spirit of the EGC paradigm.
Another aspect of \textsc{Cgal} is its generic programming approach:
algorithms access underlying data structures and primitives
through a well-defined interface, so that these layers
can be easily replaced with alternative implementations.
More specifically, we make use of the kinetic data structures
package \cite{RusselCGAL}, which provides an EGC implementation
of kinetic Delaunay triangulations in two and three dimensions.
Internally, the package contains an algebraic kernel,
providing the low-level functionality needed to handle roots of polynomials,
and a combinatorial layer,
maintaining the data structure and the certificates over time.
As mentioned earlier, we have extended
the combinatorial layer to maintaining an alpha complex.

\begin{figure}[h]
 \vspace{-0.1in} \centering
   \includegraphics[width=4cm]{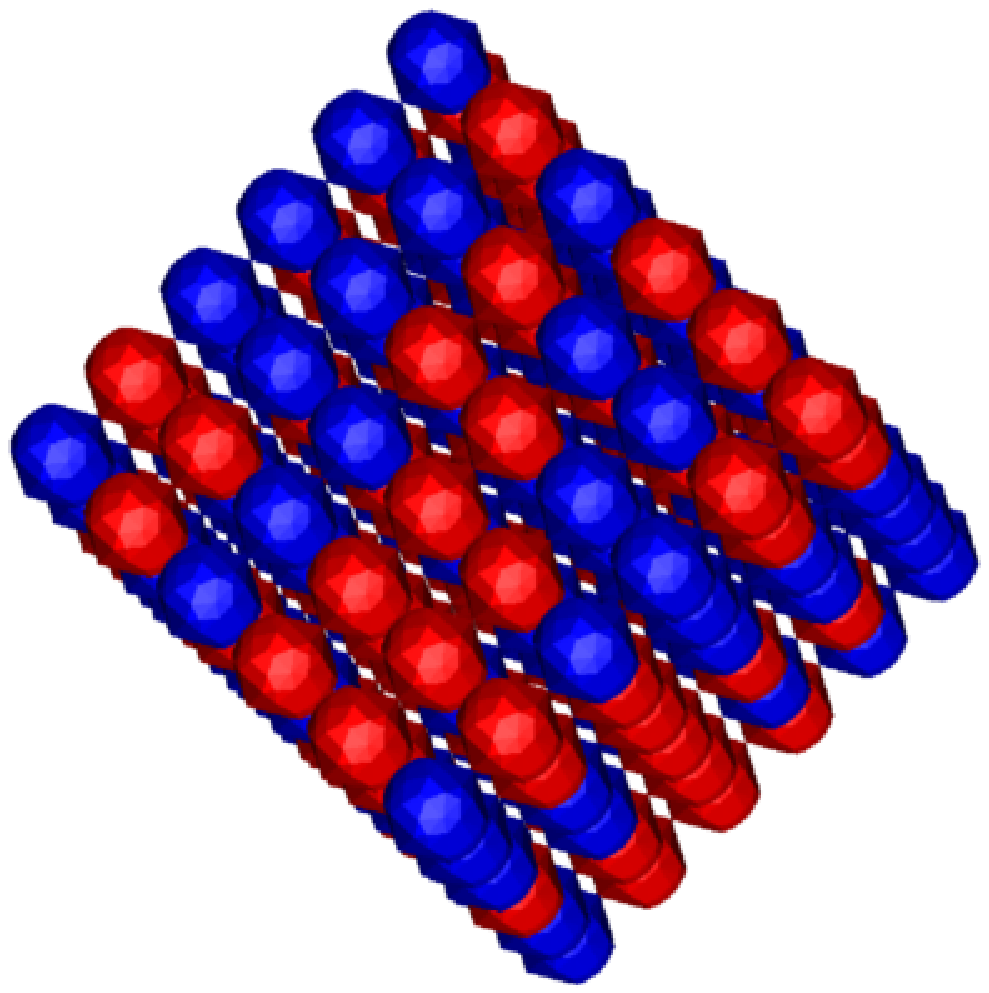}
   \includegraphics[width=4cm]{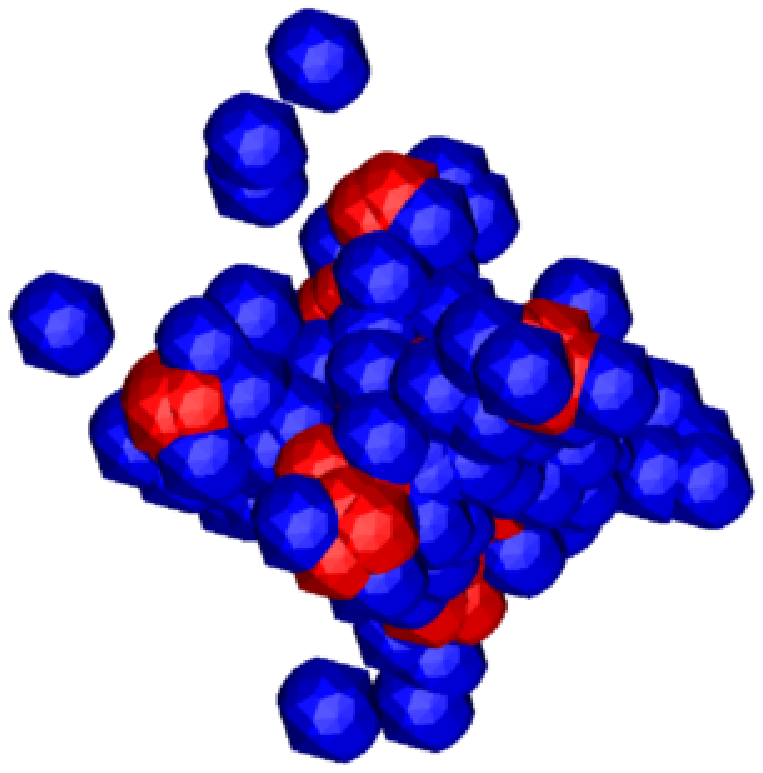}
   \includegraphics[width=4cm]{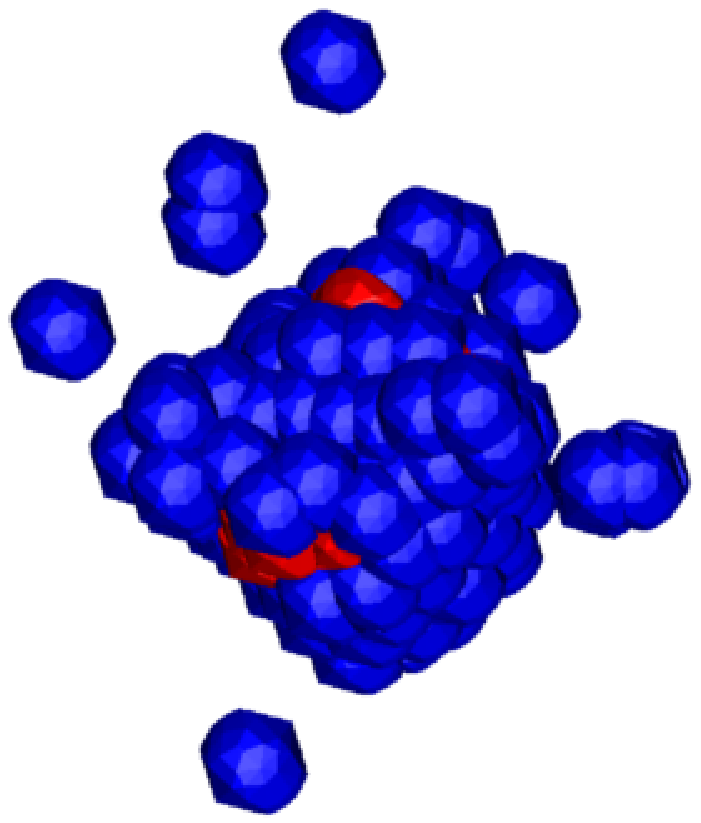}
   \includegraphics[width=4cm]{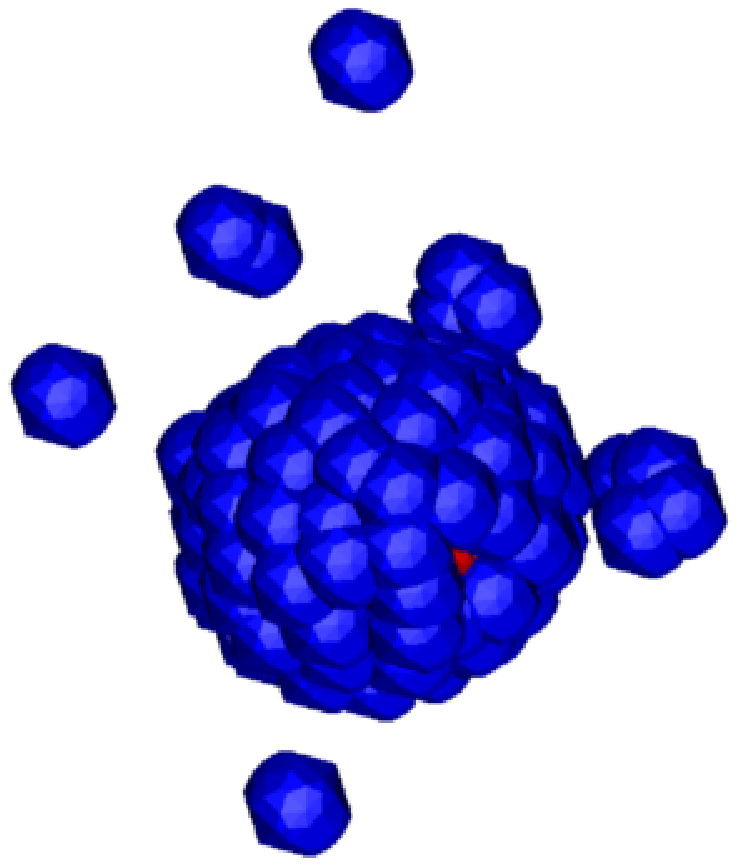}
 \vspace{-0.2in}
 \caption{The restricted Voronoi tessellation at four moments in time.
   At the beginning, the cells form a cubical grid (upper-left). 
   The cells move toward the center of the available space,
   and the blue cells begin to engulf the red cells
   (upper-right and lower-left), allowing for satellites while this happens.
   Finally, the blue cells form a sphere
   surrounding a ball of red cells (lower-right).}
 \label{fig:Vor-tessel}
\end{figure}

\paragraph{Experimental set-up.}
We use datasets obtained with the
\textsc{CompuCell3D} software\footnote{\texttt{www.compucell3d.org/}.},
which allows for the simulation of a $3$-dimensional cell segregation process
using a Monte-Carlo algorithm for energy minimization;
see the companion paper \cite{EHKK12} for more details.
We focus on simulated as opposed to observed data
because they offer a better control of the input size
and the direct accessibility of the cell trajectories.
In our particular example, the cells are colored blue or red,
each color with probability one half,
and the parameters of the simulation are chosen so that
the blue cells eventually engulf the red ones;
see Figure \ref{fig:Vor-tessel} for an illustration.
We created datasets for several input sizes.
In all cases, each trajectory represents the path of a cell nucleus
which exists throughout the entire process.
Hence, no new cells are ever inserted after the start of the process,
and no old cells are deleted before the end of the process.
The trajectories follow a global rhythm in which each trajectory
starts a new segment at each value in a common sequence of bending events.
In between two bending events, each trajectory is linear.
All experiments are performed on a Intel Core 2 Dual CPU clocked
with 2.4 GHz each, with 3 MB of cache size, and 4 GB of total memory.
The code runs under Debian Squeeze,
compiled with gcc-4.4.5 and \textsc{Cgal} version 3.9.

\begin{table}[htb]
 \centering
 {\footnotesize
  \begin{tabular}{rr|rrr|rr}
    \multicolumn{2}{c|}{\sc Input \#s}       &
      \multicolumn{3}{c|}{\sc Time in sec}     &
      \multicolumn{2}{c}{\sc \#Events}         \\ 
    \multicolumn{1}{c}{traj}                 &
      \multicolumn{1}{c|}{bends}               &
      \multicolumn{1}{c}{Del}                  &
      \multicolumn{1}{c}{Alpha}                &
      \multicolumn{1}{c|}{Medusa}              &
      \multicolumn{1}{c}{flips}                &
      \multicolumn{1}{c}{rad}                  \\  \hline
        20 &  20 &   7 &   740 &   743 &   512 &   631 \\
        20 &  40 &  29 & 1,550 & 1,553 & 1,011 & 1,335 \\
        20 &  80 &  81 & 3,205 & 3,203 & 2,019 & 2,503 \\
        20 & 160 & 188 & 6,473 & 6,484 & 3,978 & 4,506 \\ \hline
        10 &  40 &   7 &   487 &   491 &   369 &   554 \\
        20 &  40 &  29 & 1,549 & 1,556 & 1,011 & 1,335 \\
        40 &  40 &  79 & 3,975 & 3,985 & 2,874 & 2,171 \\
        80 &  40 & 229 & 9,897 & 9,904 & 7,856 & 4,977 \\
       160 &  40 & 495 &21,516 &21,741 &17,667 & 5,998
  \end{tabular}
 }
\caption{Columns from left to right:  the number of trajectories and
  bending events per trajectory,
  the time to maintain the Delaunay triangulation and the alpha complex,
  the time to compute the alpha medusa (which includes
  the maintenance of the alpha complex),
  and the number of flip and radius events.}
\label{tbl:del_alpha_compare}
\end{table}
In a first test, we compare the running times
for maintaining the Delaunay triangulation and the alpha complex,
and for constructing the alpha medusa,
all for the same datasets; see Table \ref{tbl:del_alpha_compare}.
We observe that the overhead of computing the medusa is negligible.
For this reason, we concentrate on the maintenance of the alpha complex.
Comparing the third and fourth columns of the table,
we see that the radius events slow down the algorithm by more than
a magnitude, in spite of the fact that their number
is not much larger than the number of flip events.
In the remainder of this section, we explain improvements of our
algorithm aimed at reducing the performance gap between Delaunay
and alpha complexes.

\paragraph{Number of certificates.}
The bottleneck is the construction of radius certificates
and the computation of their real roots.
Recall that in our original formulation, we maintain
a radius certificate for each edge, triangle, and tetrahedron.
Our first optimization is based on the observation that many of these
certificates are not necessary:
if a simplex is short, then all its faces are short,
and if a simplex is non-short, then all its cofaces are non-short.
\begin{result}[Optimization 1]
  Whenever a triangle or tetrahedron becomes short,
  we remove the radius certificates of its proper faces,
  and when an edge or triangle becomes non-short,
  we remove the radius certificates of its proper cofaces.
\end{result}
Of course, this implies that we sometimes have to construct certificates
that would otherwise still exist.
For example, we construct the certificate of a triangle
at the time its third edge becomes short.
On the other hand, we avoid unnecessary certificates, for
instance the certificates of the boundary edges of 
a triangle that stays short for the whole simulation.
As we see in Table \ref{tbl:reduced_cert},
the strategy saves time in practice.
We observe that the constructions of radius certificates
and the running time both decrease roughly by a factor of two.
\begin{table}[htb]
 \centering
 {\footnotesize
  \begin{tabular}{rr|rr|rr}
    \multicolumn{2}{c|}{\sc Input \#s}        &
      \multicolumn{2}{c|}{\sc Time in sec}      &
      \multicolumn{2}{c}{\sc \#Certificates}    \\
    \multicolumn{1}{c}{traj}                  &
      \multicolumn{1}{c|}{bends}                &
      \multicolumn{1}{c}{before}                &
      \multicolumn{1}{c|}{after}                &
      \multicolumn{1}{c}{before}                &
      \multicolumn{1}{c}{after}                 \\ \hline
        20 &  20 &   740 &   361 & 20,211 & 10,262 \\
        20 &  40 & 1,550 &   770 & 40,897 & 20,622 \\
        20 &  80 & 3,205 & 1,579 & 82,287 & 41,037 \\
        20 & 160 & 6,473 & 3,142 &163,511 & 80,489 \\ \hline
        10 &  40 &   487 &   248 & 12,932 &  6,324 \\
        20 &  40 & 1,549 &   770 & 40,897 & 20,622 \\
        40 &  40 & 3,975 & 1,892 &105,754 & 54,426 \\
        80 &  40 & 9,897 & 4,882 &259,848 &139,816 \\
       160 &  40 &21,516 &10,181 &566,589 &303,065
   \end{tabular}
  }
  \caption{Third and fourth columns:  the time to maintain the
     alpha complex before and after Optimization 1.
     Fifth and sixth columns:  the number of radius certificates
     before and after Optimization 1.}
\label{tbl:reduced_cert}
\end{table}

\paragraph{Degree.}
We turn to the computation of certificates.
Assuming piecewise-linear trajectories, the radius certificate
of an edge is a polynomial of degree $2$; compare with Section \ref{sec3}.
There is a standard construction of a radius certificate of a tetrahedron,
which is a polynomial of degree $8$; see \cite{WeiWeb}.
Our interest lies in the remaining triangle case.
The current \textsc{Cgal} implementation computes the squared
circumradius of a triangle $\Delta$ in $\Rspace^3$
with an expression of the form
\begin{eqnarray}
  r_\Delta^2  &=&  \frac{\mathrm{num}_x^2 + \mathrm{num}_y^2 + \mathrm{num}_z^2}
                   {4\mathrm{den}^2},
  \label{eqn:radius-old}
\end{eqnarray}
where $\mathrm{den}$ is the determinant of the matrix in \eqref{eqn:flipB},
and $\mathrm{num}_x$, $\mathrm{num}_y$, $\mathrm{num}_z$ are expressions
formed by minors of this matrix.
The corresponding certificate,
$$
  \mathrm{num}_x^2 + \mathrm{num}_y^2 + \mathrm{num}_z^2
    - 4 \alpha_0^2 \mathrm{den}^2 ,
$$
is a polynomial whose degree is $10$, which is higher than the degree
for the tetrahedron.
We replace \eqref{eqn:radius-old} by a simpler expression.
Writing $u$, $v$, $w$ for the three vertices of the triangle,
the circumradius can also be written as
\begin{eqnarray}
  r_\Delta  &=&  \frac{\Edist{u}{v} \cdot \Edist{u}{w} \cdot \Edist{v}{w}}
                 {2 \|(u-w) \times (v-w)\|} ,
  \label{eqn:radius-new}
\end{eqnarray}
a formula that is straightforward to derive using elementary matrix calculus;
see also Wikipedia \cite{wiki}.

\begin{result}[Optimization 2]
  Monitor the radius of a triangle using the
  following certificate function:
  $$
    \Edist{u}{w}^2 \Edist{u}{w}^2 \Edist{v}{w}^2
      - 4 \alpha_0^2 \|(u-w) \times (v-w)\|^2 .
  $$
\end{result}
The degree of this certificate is $6$.
We see the effect of this improvement in Table \ref{tbl:reduced_degree}.
The running time improves by more than a factor of two.
\begin{table}[htb]
 \centering
 {\footnotesize
  \begin{tabular}{rr|rr}
    \multicolumn{2}{c|}{\sc Input \#s}        &
      \multicolumn{2}{c}{\sc Time in sec}       \\
    \multicolumn{1}{c}{traj}                  &
      \multicolumn{1}{c|}{bends}                &
      \multicolumn{1}{c}{deg 10}                &
      \multicolumn{1}{c}{deg 6}                 \\ \hline
        20 &  20 &   361 &  151 \\
        20 &  40 &   770 &  342 \\
        20 &  80 & 1,579 &  734 \\
        20 & 160 & 3,142 &1,515 \\ \hline
        10 &  40 &   248 &  105 \\
        20 &  40 &   770 &  344 \\
        40 &  40 & 1,892 &  912 \\
        80 &  40 & 4,882 &2,374 \\
       160 &  40 &10,181 &5,157
  \end{tabular}
  }
  \caption{Timings for maintaining the alpha complex using
    a degree $10$ versus a degree $6$ certificate function
    for monitoring the circumradii of triangles.}
  \label{tbl:reduced_degree}
\end{table}

\paragraph{Algebraic kernel.}
As already mentioned, the \textsc{Cgal} package for kinetic data structures
contains an internal algebraic kernel, which, among other things,
is used to isolate the roots of polynomials and sort them in the event queue.
By the generic design of the package, the combinatorial layer communicates
with the kernel via a small and well-defined interface,
which makes it possible to replace the algebraic kernel
with a different implementation. 
\begin{table}[htb]
 \centering
 {\footnotesize
  \begin{tabular}{rr|rrrr}
    \multicolumn{2}{c|}{\sc Input \#s}             &
      \multicolumn{4}{c}{\sc Time in sec}            \\
    \multicolumn{1}{c}{traj}                       &
      \multicolumn{1}{c|}{bends}                     &
      \multicolumn{1}{c}{\tt kds}                    &
      \multicolumn{1}{c}{\tt ak\_d}                  &
      \multicolumn{1}{c}{filter}                     &
      \multicolumn{1}{c}{cache}                      \\ \hline
        20 &  20 &  151 &    84 &    72 &    47 \\
        20 &  40 &  342 &   176 &   152 &    98 \\
        20 &  80 &  734 &   364 &   310 &   198 \\
        20 & 160 &1,515 &   731 &   622 &   390 \\ \hline
        10 &  40 &  105 &    55 &    49 &    30 \\
        20 &  40 &  344 &   177 &   152 &   100 \\
        40 &  40 &  912 &   458 &   392 &   256 \\
        80 &  40 &2,374 & 1,180 & 1,024 &   689 \\
       160 &  40 &5,157 & 2,566 & 2,256 & 1,481
  \end{tabular}
  }
  \caption{Timings for maintaining an alpha complex
    using the {\tt kds} kernel, the {\tt ak\_d} kernel,
    the latter with Descartes filtering,
    and in addition with enabled cache.}
\label{tbl:new_kernel}
\end{table}
In recent years, a mature and generic algebraic kernel for geometric
computations has been developed \cite{BHK10}.
It has been integrated into \textsc{Cgal} and is available
since version 3.7 under the name \texttt{Algebraic\_kernel\_d}.
We refer to it as the {\tt ak\_d} kernel.
Both the internal {\tt kds} and the {\tt ak\_d} kernels use subdivision
methods for root isolation, but they differ in the strategy for detecting
empty intervals and isolating intervals.
The {\tt kds} kernel uses Sturm theory \cite[\S 7]{Yap00},
while the {\tt ak\_d} kernel is based on Descartes' rule of sign \cite{ESY06},
which leads to a better performance in practice;
see \cite{EHKMTZ09} for a comparison of various root solvers.
The difference between the third and fourth columns
in Table \ref{tbl:new_kernel} shows that exchanging the kernel
yields another improvement of roughly a factor of two.

\paragraph{Filter and cache.}
We get further optimizations by exploiting the special structure
of our experimental setup.
For any certificate, we are only interested in the roots between the current
time and the next bending event, when the certificate becomes invalid.
Many certificates do not have roots in this interval,
but may have roots outside.
The current implementation first computes all real roots
and thereafter discards the ones that lie outside the mentioned interval.
\begin{result}[Optimization 3]
  We use Descartes' rule of sign to certify the non-existence
  of roots in the interval until the next bending event,
  and if successful, we skip the root isolation algorithm.
\end{result}
The fifth column of Table \ref{tbl:new_kernel} shows the improvement.
More than ninety percent of the certificates
that do not have a root before the next bending event
are filtered out.
As a final improvement, we avoid isolating the roots of the same
polynomial multiple times.
\begin{result}[Optimization 4]
  We store polynomials together with their real roots in the
  interval until the next bending event in cache,
  which is cleared at the next bending event.
\end{result}
We see in the sixth column of Table \ref{tbl:new_kernel}
that the cache yields another substantial speed-up,
which suggests that certificates are frequently devalidated
and revalidated during the runtime of the algorithm. 
We remark that also the {\tt kds} kernel would benefit from caching.
Comparing the running times for maintaining the alpha complex
before and after the four steps of optimization,
we see that the performance improves by roughly a factor of $15$.
Moreover, compared to maintaining the Delaunay triangulation,
the optimized algorithm is slower by a factor up to $4$.
It is no surprise that the extension to alpha complexes is expensive.
After all, it requires additional radius certificates,
which have higher degrees than the flip certificates needed to
maintain the Delaunay triangulation.
We have demonstrated that with some algorithmic engineering,
the overhead needed for alpha complexes can be kept within a moderate bound.

%%%%%%%%%%%%%%%%%%%%%%%%%%%%%%%%%%%%%%%%%%%%%%%%%%%%%%%%%%%%%%%%%%%%%%%%%%
%%%%%%%%%%%%%%%%%%%%%%%%%%%%%%%%%%%%%%%%%%%%%%%%%%%%%%%%%%%%%%%%%%%%%%%%%%
\section{Discussion}
\label{sec6}
%%%%%%%%%%%%%%%%%%%%%%%%%%%%%%%%%%%%%%%%%%%%%%%%%%%%%%%%%%%%%%%%%%%%%%%%%%
%%%%%%%%%%%%%%%%%%%%%%%%%%%%%%%%%%%%%%%%%%%%%%%%%%%%%%%%%%%%%%%%%%%%%%%%%%

The main contributions of this paper are
a kinetic algorithm for alpha complexes,
its use to construct a space-time representation of a spatial sorting process
--- called a medusa ---
and the implementation of a moderately fast but correct software.
In a companion paper \cite{EHKK12}, we have demonstrated
that the theory of persistent homology applied to the time function
on the medusa quantifies the sorting process by measuring
its topological features.
While the work in this and the companion papers is limited to
simulated data, an application of our methods to observed biological data
is under way.
There is no theoretical obstacle to generalizing our algorithm and its
implementation to the weighted case, in which different Voronoi regions
are restricted to within different size balls.
A more challenging problem is the restriction to within
bodies different than balls,
e.g.\ arbitrarily oriented ellipsoids.

The first author of this paper made a considerable effort
to accelerate the implementation of the kinetic alpha complex algorithm,
since this was necessary to compute
examples of reasonable size in acceptable time;
the instances computed in \cite{EHKK12} each took about 4 hours
with our best configuration.
Nevertheless, there are opportunities to further speed up the software,
in particular on the level of the algebraic kernel.
For example, it would be desirable to restrict the root isolation
method to within a given interval, without wasting any time on roots
outside this interval.
Similarly, we expect an improvement from implementing the filter
for ruling out empty intervals in certified approximate arithmetic.
We believe that kinetic data structures are an important tool
in the topological analysis of time-varying shapes.
We hope that our work on cell segregation initiates further work on such data. 
To facilitate this research,
it would be useful if our extension of \textsc{Cgal}'s package
on kinetic data structures is transformed from an experimental branch
to a redesign of the package.
It is desirable that such a redesign solves the problem of degeneracies
in the implementation of kinetic Delaunay triangulations and alpha complexes.
Except for some special cases, the current versions of both algorithms
are not guaranteed to work correctly when two or more events happen
at exactly the same time.

\subsection*{Acknowledgments}
{\small The authors thank Viktoriia Sharmanska for discussions and
help with a $2$-dimensional prototype of our implementation.}

%%%%%%%%%%%%%%%%%%%%%%%%%%%%%%%%%%%%%%%%%%%%%%%%%%%%%%%%%%%%%%%%%%%%%%%%%%
%%%%%%%%%%%%%%%%%%%%%%%%%%%%%%%%%%%%%%%%%%%%%%%%%%%%%%%%%%%%%%%%%%%%%%%%%%

\newpage
%%%%%%%%%%%%%%%%%%%%%%%%%%%

\ignore{
\newpage
\paragraph{To do or think about.}
\begin{itemize}\denselist
  \item  In Section 5:
    \begin{itemize}
      \item  Michael, we say that only ninety percent of the root-empty
        intervals are detected.  Why not hundred percent?  Is there an
        obvious reason?
    \end{itemize}
  \item  In Section 6:
    \begin{itemize}
      \item  It seems that SoS is the right tool to make the final step
        and make the algorithm correct for not necessarily distinct events.
        Of course this is work but it seems like something that should be
        done some time in the future.  Not sure by whom.
    \end{itemize}
\end{itemize}
}

\end{document}